\documentclass[11pt,twocolumn,aps,prb,reprint,superscriptaddress]{revtex4-1}
\usepackage{pst-node,graphicx,amsmath,amssymb,braket}

\usepackage[colorlinks=true,linkcolor=blue]{hyperref}

\begin{document}

\title{Edge diffraction and plasmon launching in two-dimensional electron systems}




\author{Egor Nikulin}
\affiliation{Laboratory of 2d Materials for Optoelectronics, Moscow Institute of Physics and Technology, Dolgoprudny 141700, Russia}
\affiliation{Physics Department, M.V. Lomonosov Moscow State University, Moscow, Russia}

\author{Denis Bandurin}
\affiliation{Department of Physics, Massachusetts Institute of Technology, Cambridge, Massachusetts 02139, USA}
\affiliation{Laboratory of 2d Materials for Optoelectronics, Moscow Institute of Physics and Technology, Dolgoprudny 141700, Russia}

\author{Dmitry Svintsov}
\affiliation{Laboratory of 2d Materials for Optoelectronics, Moscow Institute of Physics and Technology, Dolgoprudny 141700, Russia}
\email{svintcov.da@mipt.ru}

\begin{abstract}
Diffraction of light at lateral inhomogenities is a central process in the near-field studies of nanoscale phenomena, especially the propagation of surface waves. Theoretical description of this process is extremely challenging due to breakdown of plane-wave methods. Here, we present and analyze an exact solution for electromagnetic wave diffraction at the linear junction between two-dimensional electron systems (2DES) with dissimilar surface conductivities. The field at the junction is a combination of three components with different spatial structure: free-field component, non-resonant edge component, and surface plasmon-polariton (SPP). We find closed-form expressions for efficiency of photon-to-plasmon conversion by the edge being the ratio of electric fields in SPP and incident wave. Particularly, the conversion efficiency can considerably exceed unity for the contact between metal and 2DES with large impedance. Our findings can be considered as a first step toward {\it quantitative} near-field microscopy of inhomogeneous systems and polaritonic interferometry.
\end{abstract}

\maketitle

\section{Introduction}
The two main experimental problems handled with near-field microscopy are determination of electromagnetic properties with sub-wavelength lateral resolution~\cite{S-SNOM,A-SNOM,Super-resolution-NFSNOM} and studies of surface wave propagation~\cite{Near-field-microscopy-SPPs,Basov_limits_plasmonics,Kopens_plasmons}. In both problems, diffraction of light at lateral inhomogenities of electromagnetic properties plays a central role. Theoretical description of such diffraction problems encounters considerable difficulty. It is associated with broken translation symmetry and inefficiency of field expansion in a plane-wave basis. For this reason, a large number of works on 'quantitative near-field microscopy'~\cite{Quant_NFM_1,Quant_NFM_2,govyadinov2013quantitative} focused on diffraction in laterally uniform layered structures. The most interesting case of laterally non-uniform systems was approached only with numerical simulations~\cite{Spatial_conductivity_patterns,Sheet_and_edge_plasmons}.

\begin{figure}[ht!]
	\centering
	\includegraphics[width=0.9\linewidth]{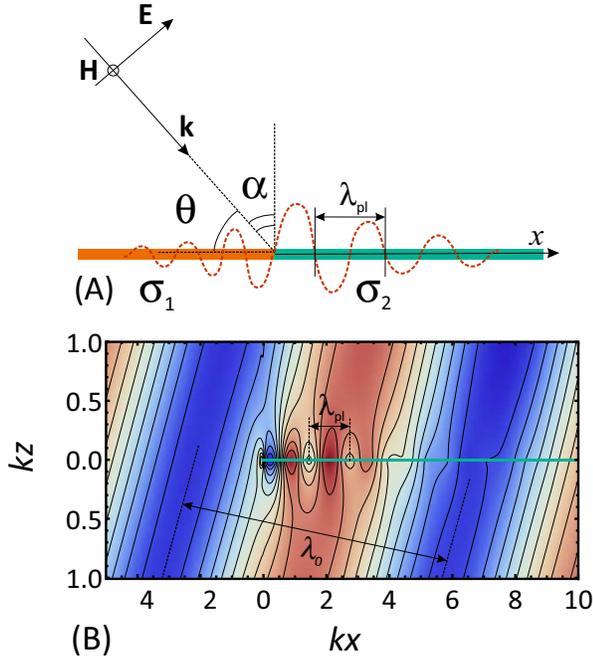}
	\caption{Geometry of the diffraction problem (A) $p$-polarised electromagnetic wave with wavelength $\lambda_0 = 2\pi/k$ is incident on the junction of 2d materials with conductivities $\sigma_1$ and $\sigma_2$ at angle $\alpha = \pi/2 - \theta$. Among other diffracted fields, it generates surface plasmon polaritons with wavelength $\lambda_{pl}$ running away from the edge (B) Example of calculated field ${\rm Im} E_x(x,z)$ for scattering at terminated ($\sigma_1 = 0$) 2DES with conductivity $2\pi\sigma_2/c = 0.2 i + 0.02$, $\theta=\pi/4$ }
	\label{fig:structure}
\end{figure}

In this paper, we present and analyze an exact solution for plane electromagnetic wave diffraction at a linear junction of materials with dissimilar surface conductivities $\sigma_1$ and $\sigma_2$ (Fig.~\ref{fig:structure} A). Such a junction can mimic a lateral contrast of local electromagnetic properties that is commonly studied in near-field experiments. At the same time, this junction is a central object in experiments with surface plasmon-polaritons (SPPs) in two-dimensional electron systems (2DES)~\cite{Basov_limits_plasmonics,Spatial_conductivity_patterns,Koppens_2p_phase_control}. A terminated edge of 2DES ($\sigma_2 = 0$) is commonly  used as a mirror for plasmons~\cite{Kopens_plasmons,Sheet_and_edge_plasmons}. It enables one to observe interference of tip-launched and edge-reflected SPPs, and to extract their dispersion and damping. A contact between 2DES and metal ($\sigma_2 = \infty$) is also commonly used as a coupler between free-space photons and SPPs~\cite{Basov_limits_plasmonics,Spatial_conductivity_patterns,Koppens_2p_phase_control}. Remarkably, such coupler can compensate arbitrarily large momentum mismatch between light and SPPs as soon as metal layer can be considered as infinitely thin.

Our exact analytical treatment reveals the spatial structure of field near the edge. It has three components with different length scales: free-field contribution repeating the periodicity of incident wave; edge diffraction contribution demonstrating oscillatory decay at the length order of free-space wavelength $\lambda_0$, and 2D SPP with wavelength $\lambda_{\rm pl}$. Interference between these three components produces decaying oscillatory fringes closely resembling those recorded in near-field studies 2D materials~\cite{Basov_limits_plasmonics,Spatial_conductivity_patterns}. Their period may considerably differ from plasmon wavelength, especially if the velocity of plasmon is close to the speed of light.

We find that the amplitude coefficient of photon-to-plasmon conversion being the ratio of SPP electric field and that in an incident wave, depends on two dimensionless parameters: angle of incidence $\alpha$ and product of 2DES conductivity and free-space impedance $\eta = 2\pi \sigma/c$. For terminated edge of 2DES, the conversion efficiency is a growing function of $(\eta'')^{-1}$ saturating at $\eta \rightarrow 0$ to a universal value $\sqrt{2}\cos\alpha$. For metal-contacted 2DES, the conversion efficiency grows as $(\eta'')^{-1/2}$ with reduction of 2DES conductivity, and can greatly exceed unity.

Our solution is based on the Wiener-Hopf technique widely applied for wave scattering at (semi)bounded objects~\cite{Noble1958MethodsBO,Valnstein,daniele2014wiener}. Our formal steps closely follow and exact solution for diffraction at a conductive metal sheet obtained more than half a century ago~\cite{Senior}. Yet one should keep in mind that the seminal work~\cite{Senior} dealt with diffraction in a physically different system, which is a terminated layer of metal being thick compared to skin depth $l_{sk}$ but thin compared to $\lambda_0$. Such metal film is modelled by surface impedance (Leontovich) boundary condition~\cite{leontovich_conditions} at each side. The true 2d electron systems are thin compared to both $l_{sk}$ and $\lambda_0$, and modelled by a two-sided boundary condition relating the discontinuity of magnetic field and surface current. However, when it comes to formulation of scattering equations, both problems become mathematically equivalent~\cite{Senior_conductive_resistive}. 

We manage to find closed-form expressions for the Fourier components of diffracted field, which makes a strong difference from Ref.~\cite{Senior} where the fields were given by hardly tractable integrals. We pay special attention to the aspects being interesting to 'plasmonic community', namely, to plasmon launching efficiency. There exists only one study~\cite{Margetis_Wiener_Hopf} of diffraction by the edge of 2DES using an analytical technique similar to Ref.~\cite{Senior}, but it was limited to the cases $\sigma_2 = 0$, $\sigma_1/c \ll 1 $, and the efficiency of plasmon launching was not analyzed therein.

\section{Exact solution of the diffraction problem}
We proceed to an exact solution for scattering of plane electromagnetic wave incident at angle $\theta = \pi/2 - \alpha$ (Fig.~\ref{fig:structure}) on the junction of 2DES with conductivities $\sigma_1$ and $\sigma_2$. The harmonic time dependence of all quantities $e^{-i \omega t}$ is as assumed and will be skipped. The magnetic field of the wave is directed along the junction of 2DES. The particular choice of polarization is dictated by possibility of TM plasmon excitation in materials with Drude-like conductivity, ${\rm Im}\sigma >0$~\footnote{TE plasmons can be excited in materials with ${\rm Im}\sigma <0$; physically such type of conductivity is realized for dominant interband transitions~\cite{Mikhailov_new_mode}. This case is left for further studies.}. 

Integral equation for scattering is most simply derived from fundamental solution of wave equation for scalar and vector potentials in Lorentz gauge (Appendix A). Physically, such equation states that net electric field ${\bf E}({\bf r})$ is the sum of incident field ${\bf E}_{0} e^{i {\bf k r}}$ and the field due to induced currents in conductors. 'Induced' fields throughout all space are governed by currents in 2DES solely, therefore it is sufficient to consider to write the scattering equation only for $z=0$ and only for $x$-component of the field:

\begin{multline}
\label{Ex1}
E_{x}(x)=E_{0} \sin \theta e^{i k_x x}-\\ - \frac{\pi}{k c}\left\{k^2+\partial_{x}^{2}\right\}\int_{-\infty}^{\infty} j_{x}\left(x^{\prime}\right) H_0(k |x-x'|) d x^{\prime},
\end{multline}
where $H_0$ is the Hankel function of the first kind, $j_x(x)$ is the electric current density, $k=\omega/c$ is the wave number, $k_x = k\cos\theta$ is the component of wave vector along 2DES, and $c$ is the speed of light (Gaussian units will be used throughout). The closure of scattering problem is achieved by relating current $j_x$ and net field $E_x$. We adopt this relation in local form, $j_x(x) = \sigma(x) E(x)$, where $\sigma(x) = \sigma_1 \vartheta(-x) + \sigma_2 \vartheta(x)$ is the conductivity distribution in 2DES, and $\vartheta(x)$ is the Heaviside step function.

In the absence of junction, the solution of diffraction problem at 2DES with uniform conductivity $\sigma_1$ would be
\begin{equation}
E^{(1)}_{F}(x)=\frac{E_0 \sin \theta }{1+\frac{2\pi \sigma_1}{c} \sin{\theta}} e^{i k_x x}.    
\end{equation}
We will subtract this 'free-field' contribution from the net electric field for improved convergence of solution at $x\rightarrow -\infty$. To this end, we introduce a new field
\begin{equation}
E^*(x)=E_x(x,0)-E^{(1)}_{F}(x),   
\end{equation}
for which the scattering problem has the form
\begin{equation}
\begin{split}
\label{Exlo}
&E^*(x)=-\frac{1 }{2k} \left\{k^2+\partial_{x}^{2}\right\}\bigg[(\eta_2-\eta_1)\int_{0}^{\infty} E^{(1)}_{F}(x')\times \\
&\times H_0(k |x-x'|) d x^{\prime}
+ \int_{-\infty}^{+\infty} \eta(x') E ^*(x')H_0(k |x-x'|) d x^{\prime}
\bigg].
\end{split}
\end{equation}
Above, we have introduced the dimensionless 2DES conductivity $\eta(x)=2\pi\sigma(x)/c$~\footnote{In SI units, $c/2\pi$ should be replaced with free-space impedance $Z_0 = 377$ $\Omega$.}. It is instructive that external field $E_0$ in Eq.~\ref{Exlo} appears multiplied by {\it contrast} of conductivity $\eta_2 - \eta_1$. Indeed, it is the dissimilarity of junction materials that produces diffraction and near fields.

Further, we shall assume that $k$ has a small positive imaginary part $k''$:  $k=k'+i k'', \ k'>0, k''>0$. Physically, this corresponds to a weakly absorbing ambient medium. Mathematically, this ensures the convergence of solution as $x\rightarrow + \infty$.

\begin{figure}[htbp]
\centering
\includegraphics[width=0.9\linewidth]{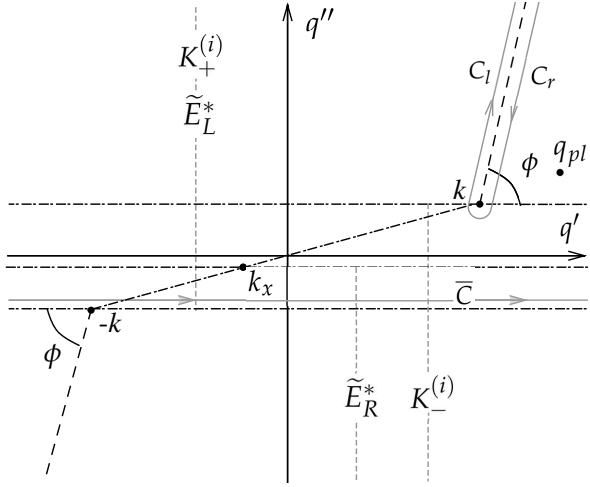}
\caption{The cut plane for the variable $q$. The cuts are represented by the black dashed half-lines. The gray dashed lines show the domains of regularity of functions appearing in the equation  \ref{Et2}. The contours of integration $C_l, C_r$ and $\overline C$ are depicted by the solid gray lines.}
\label{fig:q}
\end{figure}

Further steps follow the canonical method of Wiener and Hopf. We introduce 'left' and 'right' fields according to $E^*_L(x)=\vartheta(-x) E^*(x)$, $E^*_R(x)=\vartheta(x) E^*(x)$, and apply the Fourier transform
to the both sides of the equation \ref{Exlo}:
\begin{multline}
\label{Et1}
 \tilde E^*_L  ( q ) K^{(1)}(q) +{\tilde E^*_R  ( q )} K^{(2)}(q) =\\
 = \frac{\sqrt { k ^ { 2 } - q ^ { 2 } }}{i k}\frac{\eta_2-\eta_1 }{ k_x - q }\frac{E_0\sin\theta}{1+\eta_1 \sin\theta}.
\end{multline}
Above, $\tilde E^*_R  ( q )$ and  $\tilde E^*_L  ( q )$ are Fourier transforms of the functions $E^*_L(x)$ and $E^*_R(x)$, and
\begin{equation}
K^{(i)}(q)=1 + \eta_i \frac { \sqrt { k ^ { 2 } - q ^ { 2 } }} { k }  , \ i=1,2    
\end{equation}
is the dielectric function of inﬁnite 2d system with conductivity $\eta_i$. The functions $K^{(i)}(q)$ appear in various scattering problems involving extended 2DES, e.g. dipole radiation over a conducting plane~\cite{Margetis_dipole_over_graphene}. The zeros of $K^{(i)}(q)$ yield the dispersion of 2D plasmons $q^{(i)}_{pl}$. Once the wave vector $q$ is close to $q_{pl}$, the $q$-th Fourier component is resonantly enhanced. This fact is easily observed in scattering problems involving extended 2DES. To make a similar observation in a problem involving 2D junction, we need to make some extra steps.

First, we specify the branch cuts of $K^{(i)}(q)$ which coincide with branch cuts of the $z$-component of wave vector $k_z(q) = \sqrt{k^2 - q^2}$. We choose them to be rays starting at $q=\pm k$ and directed to infinity (see fig.\ref{fig:q}). The choice of branch for $k_z(q)$ is determined by $k_{z}(0)=k$. Once dielectric function is defined over the complex plane, we can write the solutions of $K^{(i)}(q)=0$ yielding the dispersion of 2D SPPs
\begin{equation}
    q^{(i)}_{pl} = \pm k_z ( k/\eta_i)=\pm k\sqrt{1 + (i \eta_i)^{-2}}.
\end{equation}
For passive systems with $\eta' > 0$, the wave numbers $q_{pl}$ have positive imaginary part such that $q''_{pl} > k''$.

Further solution requires splitting of functions appearing in Eq.~\ref{Et1} into those analytic in upper and lower half-planes of complex $q$-plane, and having the common strip of analyticity. It is apparent that $E_L(q)$ and $E_R(q)$ are analytic in upper (UHP) and lower (LHP) half-planes, respectively. It is now necessary to find the factorization of $K(q) = K_+(q) K_-(q)$, where $K_+$ and $K_-$ are analytic in the UHP and LHP, respectively.

The problem is solved in two steps: first, one decomposes the log-derivatives of $K_\pm(q)$ into the sum of functions analytic in UHP and LHP; second, these expressions are integrated.
According to the result of Noble \cite{Noble1958MethodsBO}, the log derivatives $d\ln K_{\pm}/dq$ can be chosen in the form 
 \begin{equation}
 \label{dKpm}
 \frac{d\ln K_{\pm}(q)}{dq}= \frac{i k }{2\eta}\left[\frac{f_{\pm}(q)-f_{\pm}(q_{pl})}{q-q_{pl}}
 +\frac{f_{\pm}(q)-f_{\pm}(-q_{pl})}{q+q_{pl}}\right], 
 \end{equation}
 where
\begin{equation}
f_{\pm}(q)=
-\frac{1}{\pi k_z(q)} \ln{\frac{\pm q-i k_z(q)}{k}}.
\end{equation}



The result of integration of (\ref{dKpm}) over $dq$ can be expressed via the dilogarithm function $\text{Li}_2(z)$:
\begin{equation}\label{new}
K_{\pm}(q)={\xi^{-\frac{1}{2}}_{\pm}} \exp \bigg[\frac{i }{{2 \pi }}\bigg(\left.\text{Li}_2\left(1+z\right)\right|^{z_1\xi_{\pm}}_{-z_1\xi_{\pm}}-\left.\text{Li}_2\left(1+z\right)\right|^{z_2\xi_{\pm}}_{-z_2\xi_{\pm}}\bigg)\bigg],
\end{equation}
where we use the notations $\left.f(x)\right|_{x_1}^{x_2}=f(x_2)-f(x_1),$ 
$$\xi_{\pm}(q)=\frac{\pm q-i k_z(q)}{k},$$
$$z_1=\xi_+(q_{pl})=i\left[1/\eta-\frac{i k_z(k/\eta)}{k}\right], \ z_2 z_1=1.$$
Here the branch cut for each of $\text{Li}_2$ function and for the square root must not intersect the branch cut of the its argument as a function of $q$, which corresponds to the branch cut for $q$  shown at fig.\ref{fig:q}. 

After the procedure of factorization, Eq.~(\ref{Et1}) takes the form
\begin{multline}
\label{Et2}
\tilde{E}^*_{L}(q) M_+(q)-\frac{C^{(1)}_{F}}{i}\frac{M_+(q)-M_+(k_x)}{k_x-q}\\
=-\tilde{E}^*_{R} M_{-}(q)+\frac{C^{(1)}_{F}}{i} \frac{M_{-}(q)+M_{+}(k_x)}{k_x -q},
\end{multline}
where $$M_+(q)=K^{(1)}_+(q)/{K^{(2)}_+(q)},\ M_-(q)=K^{(2)}_-(q)/{K^{(1)}_-(q)},$$
$$C^{(i)}_F=\frac{E_0 \sin \theta }{1+\eta_i \sin{\theta}}, \ i=1,2.$$

It can be seen that the RHS and the LHS of the  equation \ref{Et2} are regular in the lower half-plane ($\Im q<\Im(k \cos{\theta})$) and upper half-plane ($\Im q>\Im(-k)$)  correspondingly and tend uniformly to zero as $|q|\to\infty$. On applying the  Liouville's theorem, we conclude that both left and right-hand sides are identically zero. Thus, we arrive at the solution for Fourier-transformed fields:
\begin{gather}
\widetilde{E}^*_{L}(q)=-\frac{i C^{(1)}_{F}}{q-k_x} \left\{1-\frac{M_{+}(k_x)}{M_{+}(q)}\right\} ,\\
\widetilde{E}^*_{R}(q)= \frac{i C^{(1)}_{F}}{q-k_x} \left\{1-\frac{M_{+}(k_x)}{M_{-}(q)} \right\}.
\end{gather}
The inverse Fourier transform should be taken along the contour $\overline C$ being the straight line lying within the strip $\Im(-k)<\Im q<\Im (k \cos{\theta})$:
\begin{gather}\label{exfR}
\frac{{E_R } ( x )}{C^{(1)}_F}  = -\frac{i { M_{+} ( k_x ) }}{2\pi }\int\limits_{\overline C}\frac { e^{i q x} } {  q-k_x } \frac { K^{(1)} _ { - } (q)} { K^{(2)} _ { - } ( q ) } d q,\\
\label{exfL}
\frac{{E_L } ( x )}{C^{(1)}_F}  = e^{i k_x x} -\frac{i { M_{+} ( k_x) }}{2\pi }\int\limits_{\overline C}\frac { e^{i q x} } {  q-k_x} \frac { K^{(2)} _ { + } (q)} { K^{(1)} _ { + } ( q ) } dq.
\end{gather}


To provide physical insights into the structure of the field, it is instructive to derive an alternative representation for inverse Fourier transforms (\ref{exfL}) and (\ref{exfR}). For $x>0$, we can close the path of integration $\overline C$ in \ref{exfR} in the upper half-plane  and apply the residue theorem with Jordan's lemma. Within the closed contour, the integrand has a pole at the point $q=k_x$. It corresponds to the field repeating the spatial structure of incident wave. Second, it has a first-order pole at $q= q^{(2)}_{pl}$  being the zero of $K^{(2)} _ { - } (q)$ in the UHP. Physically, it corresponds to SPPs supported by the medium with conductivity $\eta_2$. ~\footnote{It is instructive that plasmon pole can disappear for particular values of conductivities $\eta_i$ and directions of branch cut. Particularly, for vertical branch cut with $\phi = \pi/2$ plasmon pole is absent if 
$|\eta|>1, 0<\arg{\eta<\pi/5}.$ . At the same time, the direction of cut can be always chosen such that the plasmon pole exists. Naturally, the solution for electric field does not depend on the presence of plasmon pole as soon as the cut does not cross the strip of analyticity. This fact is evident from Eqs.~(\ref{exfL}) and (\ref{exfR}).}

Denoting the integrand in Eq.~(\ref{exfR}) for the 'right' field by $F(q)$, we present the result of integration as 
\begin{multline}\label{res}
\int\limits_  {\overline C}F(q)d q=2 \pi i (\text{Res}[F,q_{pl}]+\text{Res}[F,k_x])\\
-\bigg(\int\limits_  {C_r}F(q)d q+\int\limits_  {C_l}F(q)d q\bigg),
\end{multline}
where $C_l, C_r$ are the left and the right lips of the branch cut, shown in the figure 2. Similar reasoning can be repeated for $E_L(x)$ at $x<0$ by closing the integration path in the LHP. However, the  pole at $q=k_x$ is now outside of the closed loop. The pole at $q = -q^{(1)}_{pl}$ may fall within the loop for certain values of conductivity. Still, as soon as we are interested in metal-contacted ($\sigma_1 = \infty$) or terminated 2DES ($\sigma_2 = 0$), the plasmon poles in the LHP are also absent.


\section{Analysis of the solution}
The above solution of diffraction problem is applicable to any junction of 2d materials with dissimilar conductivities $\sigma_1$ and $\sigma_2$. Below, we will analyze the fields for the cases of main interest in plasmonic experiments: terminated 2DES ($\sigma_1 = 0$) and metal-contacted 2DES ($\sigma_1 = \infty$). These fields will be supplied with superscripts $^{(0)}$ and $^{(\infty)}$, respectively.

\subsection{Spatial profile of the field}
The field generated upon diffraction in 2DES ($x>0$) can be represented as a sum of three components with physically transparent meaning. The first one  is the same as upon scattering at extended system, it repeats the spatial structure of external field and is most simply found from Fresnel's matching conditions. We call it 'free-field contribution'; it is given by 
\begin{equation}
    E^{0,\infty}_{F}(x) =  \frac{E_0 \sin\theta e^{i k_x x }}{1+\frac{2\pi\sigma}{c}\sin\theta}.
\end{equation}

\begin{figure}[ht!]
\centering
\includegraphics[width=0.85\linewidth]{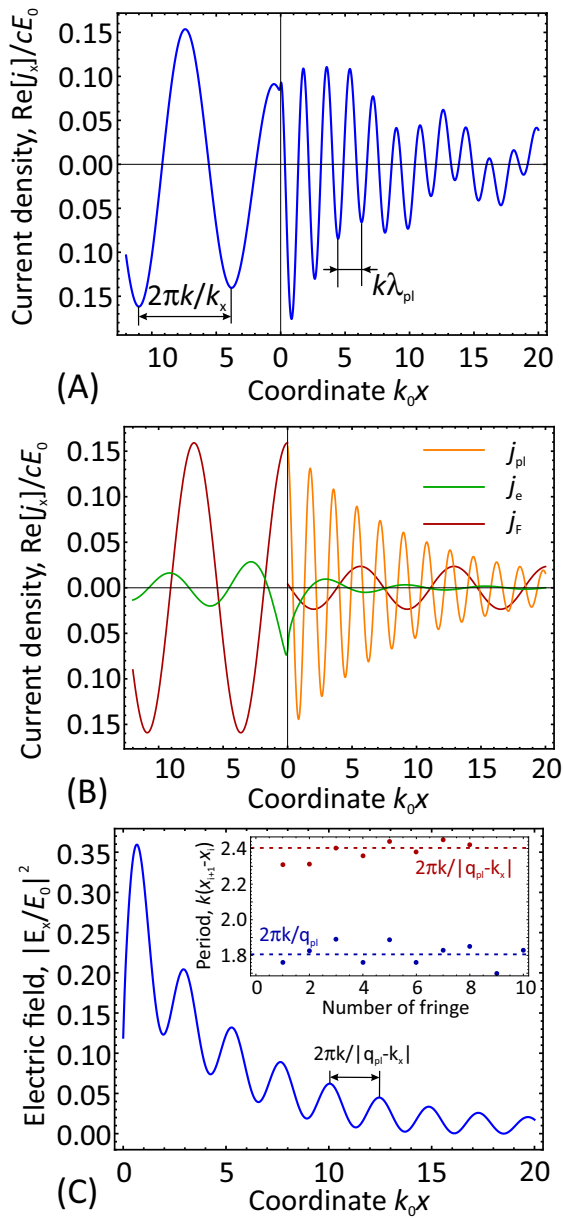}
\caption{Calculated diffraction patterns for metal ($x<0$)/2DES ($x>0$) contact in the junction plane ($z=0$): (A) Real part of electric current density (normalized by $cE_0$) (B) Expansion of the net current into Fresnel, edge, and plasmonic components (shown with red, green, and orange lines, respectively) (C) Squared amplitude of in-plane electric field. Inset shows the periods of fringes determined as spacing between maxima in (A) [blue dots] and (C) [red dots], dashed lines show the expected period of fringes from analytical considerations. Calculations were performed at $\eta_2 = 0.3 i+0.01$, $\theta = \pi/6$}
\label{fig:fields}
\end{figure}

The second one represents near-field harmonics excited non-resonantly near the edge, which we denote $E_{e}$. Mathematically, it comes from the integration along the branch cut of the Green's function, and is given by
\begin{multline}
\frac{E^{(\infty)}_{e}(x)}{E_0} = -\frac{\cos{\frac{\theta}{2}}e^{  i k x +3i\pi/4} }{\pi K_+(k_x)}\\
\times\int\limits_0^{\infty } \frac{d\tau \sqrt{2 \tau}  K_+[k(1 + i \tau)] e^{-k x \tau}}{[1+ i \eta^2 \tau(2 + i \tau)] \left[ i \tau + 2 \sin^2 \frac{\theta}{2}\right]},
  \end{multline}
 \begin{multline}  \frac{E^{(0)}_{e}(x)}{E_0}  = -\frac{ \eta \sin \theta e^{ i k x + 3i \pi/4}}{ \pi  K_+(k_x )}  \\ \times\int\limits_{0}^{+\infty} \frac{d \tau \sqrt{\tau \left(2 + i \tau\right)}  K_{+}[k(1+i \tau)] e^{- k x \tau}}{[1+ i \eta^{2} \tau (2+ i \tau )]\left[ i \tau + 2 \sin^2 \frac{\theta}{2}\right]},
 \end{multline}
Such contribution decays rapidly at distances $x \gg k^{-1}$ from the edge according to the law $e^{i k x}/(k x)^{3/2}$.

The component of our main interest is the surface plasmon-polariton generated upon edge diffraction. Mathematically, it comes from the residue of the integrand at $q = q_{pl}$. The  plasmon field oscillates in space according to $E_{pl} (x) = E_{pl}(0)e^{i q_{pl} x}$, while the wave amplitudes are given by 
\begin{gather}
\label{eq:pl_metal}
\frac{E^{(\infty)}_{pl}(0)}{E_0} = \sqrt{2} i \eta^{-2} \cos \frac{\theta}{2} \frac{K_+(q_{pl})}{K_+(k_x)} \frac{\sqrt{k^3(q_{pl}-k)}}{q_{pl}(q_{pl} - k_x)},\\
\label{eq:pl_half-plane}
\frac{E^{(0)}_{pl}(0)}{E_0} = \eta^{-2} \sin\theta \frac{K_+(q_{pl})}{K_+(k_x)} \frac{k^2}{q_{pl}(q_{pl} - k_x)}.
\end{gather}

An example of the calculated current ${\rm Re} j^{(\infty)}_x(x)$ as well as its component-wise expansion are shown in Figs.~\ref{fig:fields} (A) and (B), respectively. The short-period fringes of plasmonic component are 'threaded' on Fresnel-type current with longer period $2\pi/(k\cos\theta)$ and on decaying edge field with period $2\pi/k$.

It is remarkable that plasmonic interference pattern is formed in the profile of 'intensity' $|E_x(x)|^2$ despite there are no reflected waves to interfere with. An example of such pattern is shown in Fig.~\ref{fig:fields}(C). Actually, the interference occurs between plasmon field $E_{pl}(0)e^{iq_{pl}x}$ and other two long-period components $E_e(x)+E_F(x)$. We have verified that the main contribution to $|E_x(x)|^2$ comes from squared modulus of plasmon field $|E_{pl}(x)|^2$ and its interference with free field, $2{\rm Re}E_{pl}(x)E^*_{F}(x)$, at least in the limit $|\eta_2|\ll 1$. These observations should simplify the interpretations of near-field plasmonic experiments~\cite{Basov_limits_plasmonics}.

The spatial profile of intensity near the edge should oscillate with three characteristic wave vectors $|q_{pl} - k_x|$, $|q_{pl} - k|$, $|k - k_x|$, as there are three running waves starting at the edge. However, the decay length of $E_e(x)$ is order of oscillation period, and the interference with wave vector $k$ is inactive. It is remarkable that commonly accepted~\cite{Basov_limits_plasmonics,Spatial_conductivity_patterns} oscillation with spatial period of $\lambda_{pl}$ is, strictly speaking, absent in the intensity profile. The true spatial period is longer due to plasmon-photon interaction, and is given by $2\pi/(q_{pl}-k_x)$ [inset to Fig.~\ref{fig:fields}(C)]. On the other hand, the period extracted from complex field amplitude in Fig.~\ref{fig:fields} is uniquely related to plasmon wavelength and equals $2\pi/q_{pl}$. These facts make extraction of plasmon dispersion from mapping of near fields a non-trivial problem, especially if the conductivity of 2DES approaches the speed of light~\cite{Damping_reduction,Relativistic_Plasmon}.

Having an exact solution for edge scattering, it is instructive to comment on the frequently mentioned field enhancement at the edge. For terminated 2DES, there's no current flow into the edge, and hence $E(x\rightarrow +0) = 0$. At the same time, the dynamically-induced charges at the edge produce singular field {\it outside of 2DES},  $E(x\rightarrow -0) \propto x^{-1/2}$. In metal-contacted 2DES, the current is continuous across the edge, and approaches a universal value $c E_0/(2\pi)$ on the metal surface far away from the junction. Therefore, field at the edge of 2DES can be estimated as $E_0(c/2\pi\sigma)$. It is differs from incident field by the factor $c/2\pi\sigma$, which is above unity for 2DES with moderately low conductivity.

\subsection{Plasmon launching efficiency}
The functions (\ref{eq:pl_metal}) and (\ref{eq:pl_half-plane}) evaluated at $x=0$ can be called 'amplitude conversion efficiencies' between free-space photons and 2D SPPs and are applicable at any value of scaled 2DES conductivity $\eta = 2\pi\sigma/c$. Here, we will focus on weakly dissipative systems, $\eta'' \gg \eta'$, though the case of good conductors with 'superluminal' conductivity $\eta' > 1 \gg \eta''$ may also present considerable interest~\cite{falko_khmelnitskii}.

\begin{figure}[ht!]
\centering
\includegraphics[width=0.85\linewidth]{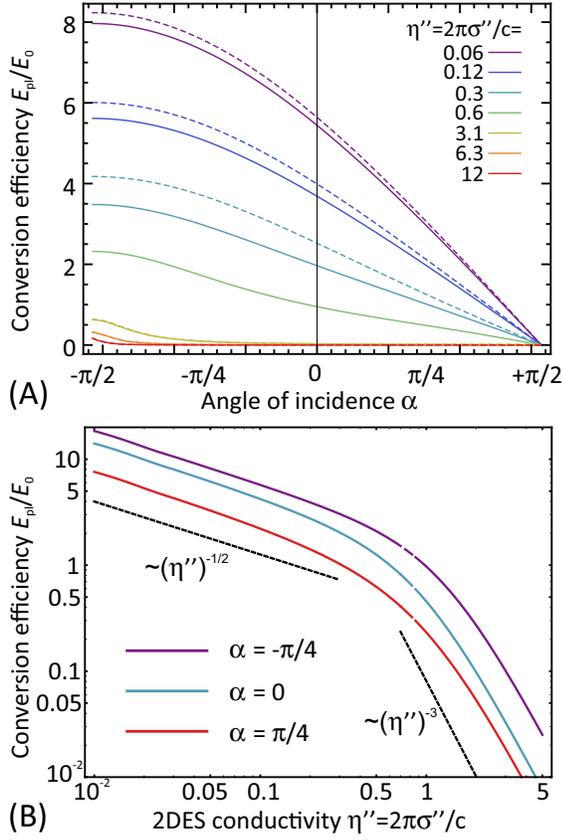}
\caption{Efficiency of photon-to-plasmon conversion upon diffraction at metal-contacted 2DES (A) vs angle of incidence at various ratios $\eta''=2\pi\sigma''/c$ (B) vs conductivity $\eta''$ at three angles of incidence. Dashed lines show analytical approximations [Eqs. (\ref{K_pl_metal_small}) and (\ref{K_pl_metal_high})] applied for cases of small and high conductivity.}
\label{fig:conv1}
\end{figure}

\begin{figure}[ht!]
\centering
\includegraphics[width=0.85\linewidth]{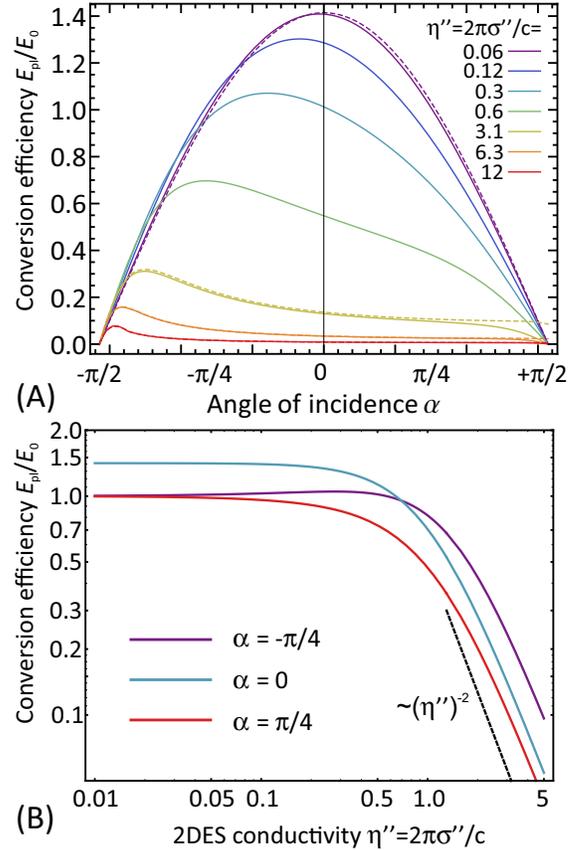}
\caption{Efficiency of photon-to-plasmon conversion upon diffraction at terminated edge of 2DES vs angle of incidence at various $\eta''$. Three lowest dashed lines show analytical approximation (\ref{K_pl_2DES_high}) valid at high conductivity, upper dashed line shows the limiting conversion efficiency $\sqrt{2}\cos\alpha$}
\label{fig:conv2}
\end{figure}

Evaluating these conversion efficiencies vs angle of incidence at different $\eta''$, we observe remarkably different patterns for metal-contacted and terminated 2DES (Figs.~\ref{fig:conv1} and \ref{fig:conv2}, respectively). The conversion at metal/2DES junction steadily grows with reduction in $\eta''$ as $(\eta'')^{-1/2}$:
\begin{equation}
\label{K_pl_metal_small}
R^{(\infty)}\approx \frac{2e^{\frac{1}{2} \cos \theta |\eta|}}{|\eta|^{1/2}} \cos\frac{\theta }{2}.
\end{equation}
This fact can be attributed to enhancement of electric field near the metal edge contacting with weakly conducting 2DES. Remarkably, the conversion efficiency can much exceed unity, as evident from both analytical expressions and calculations shown in Fig.~\ref{fig:conv1}. This does not contradict the energy conservation, as the plasmon free path also shrinks as $|\eta| \rightarrow 0$. From the other hand, it is impossible to define a dimensionless energy conversion coefficient in our scattering problem, as the extent of incident wave is infinite, while plasmon is localized at finite length.

Another interesting observation from Fig.~\ref{fig:conv1} (A) is maximization of photon/plasmon conversion coefficient at a gliding incidence. This fact is surprising due to absence of in-plane component of electric field in the incident wave. Despite that, electric current density in a perfect conductor stays at the finite value ($j \sim c H_{\rm inc}$) necessary to screen the incident field, independent of the angle. We may speculate that it is the current density in metal contact, not the incident electric field, that launches a surface wave in 2D electron system~\footnote{If metal contact has finite conductivity $\sigma_M$, the conversion efficiency would rapidly drop to zero at angles $\theta \sim \sigma_M/c$.}.

The plasmon launching by metal/2DES junction becomes inefficient at $\eta''\gg 1$ and scales as $(\eta'')^{-3}$:
\begin{equation}
\label{K_pl_metal_high}
R^{(\infty)} \approx \frac{2e^{-\frac{\theta  \csc \theta }{\pi |\eta|}}}{|\eta| (1 +   4 |\eta|^2 \sin^2 \theta/2 )}.
\end{equation}
We can speculate that metal and 2DES become electromagnetically indistinguishable in such limit, and the edge is effectively 'washed out'. 

The scaling of conversion efficiency for terminated 2DES is different. With reduction in 2DES conductivity, it approaches a universal limit of $\sqrt{2}\sin\theta$. To the next order in $|\eta|$, it reads
\begin{equation}
\label{K_pl_2DES_small}
R^{(0)}\approx\sqrt{2} \sin \theta  e^{-\frac{1}{2} \left| \eta \right|  \cos \theta }\frac{1+\left| \eta \right|  \cos \theta }{1+\frac{\left| \eta \right| ^2}{2}}.
\end{equation}
Finiteness of $R$ at nearly-zero 2DES conductivity is quite remarkable; one might expect the diffraction to be absent as the incident light meets no obstacle. In reality, both plasmon wavelength and decay length shrink at $\sigma''\rightarrow 0$, and finite plasmonic fields exist in progressively smaller domains. Therefore, diffraction vanishes at $\sigma\rightarrow 0$ {\it on average}.

In the opposite limit $\eta \gg 1$, the conversion efficiency for terminated 2DES also vanishes as $(\eta'')^{-2}$
\begin{equation}
\label{K_pl_2DES_high}
R^{(0)} \approx \frac{4 \sin \theta/2}{1 +   4 |\eta|^2 \sin^2 \theta/2  }.
\end{equation}

\section{Discussion and possible extensions}

We have presented and analyzed an exact solution for electromagnetic wave scattering at the junction of 2d electron systems with dissimilar conductivities $\sigma_1$ and $\sigma_2$. Our solution enabled the analysis of conversion between free-space electromagnetic waves and SPPs at such junction. Remarkably, this technique can be extended to a wide variety of electromagnetic problems involving 2D junctions, which we retain for further studies. Particularly, it is possible to study the scattering of surface waves by a step-like contrast of 2D conductivity or even by and edge of metal gate with uniform 2DES beneath it. Such studies are important for design of plasmon-enhanced photodetectors~\cite{bandurin2018resonant,Cai_THz_detection}, and were addressed previously only with numerical~\cite{Sydoruk_gated_ungated,Sydoruk_junctions_retardation} or approximate~\cite{SPP_2D_junction1,SPP_2D_junction2,Amplified_reflection} techniques~\footnote{The problem of plasmon scattering by conductivity contrast was solved exactly in the non-retarded limit for normal incidence in~\cite{Scattering_by_conductivity_contrast_exact}}. 

Possibly the most restrictive assumption of our analysis was the uniformity of dielectric properties of materials enclosing the junctions. Once the dielectric permittivities of upper ($\varepsilon_u$) and lower ($\varepsilon_l$) materials are dissimilar, the functions appearing in Eq.~(\ref{Et1}) should be modified according to:
\begin{gather}
  \label{E-replace}
E_0\sin\theta \rightarrow \frac{2E_0\sin\theta}{\sqrt{\frac{\varepsilon_l}{\varepsilon_u}}   \frac{\sin\theta}{\sin\theta'}+1}, \\  
\label{K-replace}
    K^{(i)}(q)\rightarrow 1 + \frac{2\eta_i}{\frac{\varepsilon_u}{\sqrt{k_u^2-q^2}}+\frac{\varepsilon_l}{\sqrt{k_l^2-q^2}}},  
\end{gather}
where $k_{u,l} = \omega \sqrt{\varepsilon_{u,l}}/c$ are the wave vectors in upper (lower) dielectric material. While the replacement (\ref{E-replace}) is readily performed in all our final results, the explicit factorization of (\ref{K-replace}) looks a challenging problem~\cite{Coblin_Pearson}. We may suggest that replacement $\eta = 2\eta/(\varepsilon_u + \varepsilon_l)$ would well describe the efficiency of plasmon launching by 2d edge residing between dissimilar dielectrics, at least in the case $|\eta| \ll 1$. Our expectation is based on the fact that dielectric functions of 2DES $K^{(i)}(q)$ depend only on average permittivity $(\varepsilon_u + \varepsilon_l)/2$ in the short-wavelength limit $q\ll k$.

The obtained structure of the diffracted field is determined fully by the angle of incidence and scaled 2DES conductivity $\eta$. The latter can be depend on frequency $\omega$, and such dependencies were studied in detail for graphene and other 2D materials, as well as for III-V based quantum wells. Irrelevance of functional dependence $\eta(\omega)$ comes from the fact that solution is performed in spatial domain, with the radiation frequency fixed. Yet, we limited ourselves to the local models of conduction where electric current $j(x)$ is proportional to electric field at the same point, $E(x)$. The presence of carrier diffusion and/or viscosity~\cite{Bandurin_viscosity} invalidates local conduction models in the 'transition layer' in immediate vicinity of the edge. We can estimate the size of this region as $\sim v_F/\omega$, where $v_F$ is the characteristic electron Fermi velocity in 2DES~\cite{Crossover}. Inclusion of non-local conduction into exact Wiener-Hopf solution technique is possible within hydrodynamic approach to electron transport~\cite{Cohen_viscosity}. 

The diffraction of obliquely incident electromagnetic wave can be also handled with Wiener-Hopf technique, though it leads to a {\it system} of Wiener-Hopf equations. Fortunately, equations of such system are readily decoupled~\cite{senior_oblique}. The problem of oblique incidence is tightly linked to the problem of edge plasmons~\cite{schmidt2014universal}, which behaviour in the presence of retardation was so far studied only with approximate methods~\cite{zabolotnykh2016edge}.

The presented solution for electromagnetic scattering by the junction of 2d electron systems can be considered as a first step toward quantitative near-field microscopy. Indeed, once our solution is generalized to the case of oblique incidence, it is straightforward to consider dipole radiation over a 2d junction. Registration of such radiation from the sub-wavelength tip is the basis for near-field microscopy.

This work was supported by the grant \# 16-19-10557 of the Russian Science Foundation. The authors thank Aleksey Nikitin for valuable discussions.

\appendix
\section{Derivation of integral equation for diffraction}

Electric field generated upon diffraction at a linear junction of 2DES is the superposition of incident field ${\bf E}_{\rm inc}({\bf r})$ and field induced by charges and currents in 2DES ${\bf E}_{\rm ind}({\bf r})$:
\begin{equation}
    {\bf E}({\bf r},t) = {\bf E}_{\rm inc}({\bf r},t) +{\bf E}_{\rm ind}({\bf r},t).
\end{equation}
Incident field represents a harmonic $p$-polarized electromagnetic wave:
\begin{gather}
    {\bf E}_{\rm inc}({\bf r},t) = {\bf E}_0 e^{i k x \cos\theta - i k z \sin\theta - i \omega t},\\
    {\bf E}_0 = \{E_0\sin\theta,0,E_0\cos\theta\}.
\end{gather}
We shall skip the harmonic time dependence $e^{-i\omega t}$ by passing to the Fourier components ${\bf E}_\omega ({\bf r})$.

Induced field ${\bf E}_{\rm ind}({\bf r})$ emerges due to currents (with bulk density ${\bf J}_{\omega}({\bf r})$ ) and charges (with bulk density $Q_{\omega}({\bf r})$ ) in 2DES. Relation between them is most easily found from fundamental solution of wave equation for the scalar $\varphi_\omega$ and vector ${\bf A}_\omega$ in the Lorentz gauge:
\begin{gather}
\label{A-potential}
{\bf A}({\bf r},t) = \frac{1}{c}\int\limits_{\rm conductor}{\frac{{\bf J}_\omega({\bf r}') e^{i k |\Delta {\bf r}|}}{|\Delta {\bf r}|} d{\bf r}'},
\\
\label{Phi_potential}
{\varphi({\bf r},t)} = \int\limits_{\rm conductor}{\frac{{Q_\omega ({\bf r}') e^{i k |\Delta {\bf r}|}}}{|\Delta {\bf r}|} d{\bf r}'}.
\end{gather}
In the above equations, the integration is performed over the volume of conductor, $\Delta {\bf r} = {\bf r}' - {\bf r}$ is the distance between observation point and location of charge/current element, $k = \omega/c$ is the wave number. The term $e^{i k |\Delta {\bf r}|}$ is responsible for the retarded nature of electromagnetic potentials.

We now use the two-dimensional nature of the conductor and pass to the sheet current and charge densities, ${\bf J}_\omega ({\bf r}) = {\bf j}_\omega(x) \delta(z)$,  $Q_\omega({\bf r}) = \rho_\omega(x)\delta(z)$. Integration in (\ref{A-potential}) and (\ref{Phi_potential}) over vertical coordinate is trivial. Integration along the edge ($y$-coordinate) can be absorbed into the definition of electromagnetic propagator (Green's function)
\begin{equation}
G (k |x-x'|) = \int\limits_{-\infty}^{\infty}  \frac{dy e^{i k [(x-x')^2+y^2]^{1/2}}}{\sqrt{(x-x')^2+y^2}}
= i \pi H_0 \left(k |x-x'| \right),    
\end{equation}
where $H_0$ is the Hankel function. In this notation, the potentials are linked to charges and currents as
\begin{gather}
\label{A-Green}
{\bf A}_\omega({\bf r}) = \frac{i\pi}{c}\int_{-\infty}^{\infty}{dx {\bf j}_\omega(x')  H_0(k |x-x'|)},\\
\label{Phi-Green}
{\varphi }_\omega({\bf r}) =i\pi \int_{-\infty}^{\infty}{dx \rho_\omega(x') H_0( k |x-x'|)}.
\end{gather}


Further, we shall express all quantities ($\bf A$, $\varphi$, $\rho$, $\bf j$) via electric field. If the $H$-field is directed along the edge ($y$-axis), all quantities can be expressed via $E_x$. First, we use the relation between $\bf E$ and retarded potentials:
\begin{equation}
\label{E-phi-A}
{\bf E}_\omega = -\frac{\partial \varphi_\omega}{\partial {\bf r}} + \frac{i \omega}{c}{\bf A}_\omega.
\end{equation}
Second, charge density is linked to current density via continuity equation
\begin{equation}
\label{Continuity}
-i\omega\rho_\omega + \frac{\partial j_{x\omega}}{\partial x} = 0.
\end{equation}
Combination of (\ref{A-Green}), (\ref{Phi-Green}), (\ref{E-phi-A}) and (\ref{Continuity}) leads us to Eq.~(\ref{Ex1}) of the main text (subscript $\omega$ is skipped):
\begin{multline}
E_{x}(x)=E_{0} \sin \theta e^{i k \cos \theta x}-\\ - \frac{\pi}{k c}\left\{k^2+\partial_{x}^{2}\right\}\int_{-\infty}^{\infty} j_{x}\left(x^{\prime}\right) H_0(k|x-x|) d x^{\prime}.
\end{multline}
We note that changing the order of taking the derivative and integration over $x'$ is possible if current density at the junction is continuous. This is indeed the physical case: discontinuity of alternating current would lead to accumulation of linear charge density at the junction. The latter could produced unphysical log-divergent potentials and infinite electric fields.
\bibliography{sample}


\end{document}